\documentclass[aps,twocolumn,prb,superscriptaddress,amsmath,tightenlines]{revtex4}

\usepackage[dvipdfmx]{graphicx}
\usepackage{epsfig}

\newcommand{\ra}{\rangle}
\newcommand{\la}{\langle}

\newcommand{\fc}{\frac}

\begin{document}

\title{Quantum Annealing Machines Based on Semiconductor Nanostructures}
\author{Tetsufumi Tanamoto}

\author{Yoshifumi Nishi}

\affiliation{Corporate R \& D center, Toshiba Corporation,
Saiwai-ku, Kawasaki 212-8582, Japan}

\author{Jun Deguchi}
\affiliation{Institute of Memory Technology R \& D, Toshiba Memory Corporation, 
Saiwai-ku, Kawasaki 212-8520, Japan} 

\begin{abstract}
The development of quantum annealing machines (QAMs) based on superconducting qubits 
has progressed greatly in recent years and these machines are now widely used in both academia and commerce.
On the other hand, QAMs based on semiconductor nanostructures 
such as quantum dots (QDs) appear to be still at the initial elementary research stage
because of difficulty in controlling interaction between qubits. 
In this paper, we review a QAM based on a semiconductor 
nanostructures such as floating gates (FGs) or QDs from the viewpoint of the integration of qubits.
We theoretically propose the use of conventional high-density memories such as NAND flash memories
for the QAM rather than construction of a semiconductor qubit system from scratch.
A large qubit system will be obtainable 
as a natural extension of the miniaturization of commercial-grade electronics, 
although further effort will likely be required to achieve high-quality qubits.
\end{abstract}


\maketitle
\section{Introduction}
Since the commercial success of D-Wave's superconducting machines~\cite{Dwave,Dwave2}, 
development of the quantum annealing machines (QAMs) has been one of the hottest 
topics in science and technology.
The theoretical background can be traced back to Nishimori’s work in 
the 1990s~\cite{Nishimori,Finnila}. 
The progress of research on QAMs has 
also stimulated new investigations on annealing methods based on digital computers~\cite{Yamaoka, Fujitsu}.
QAMs are expected to solve the combinatorial
optimization algorithms of NP-hardness problems in a shorter time than 
classical annealing methods. 
Solvers of this type are required for artificial intelligence (AI), 
whose progress is a momentous trend 
and expected to bring about drastic change in society.
Faster solving of combinatorial optimization problems by QAMs
has the potential to lead to more efficient development of AI algorithms.  
QAMs have also been used recently to investigate 
the quantum Boltzmann machine\cite{Dumoulin,Kieferova,Benedetti}.

Many combinatorial problems, including the traveling salesman problem,
can be mapped to the problems to find ground states of the Ising Hamiltonian,
 expressed by~\cite{Lucas,MAXCUT,Kahruman}
\begin{equation}
H=\sum_{i<j} J_{ij} s_i^z s_j^z 
+ \sum_i h_i s_i^z, 
\end{equation}
where the variable $s_i$ is a classical bit of two values($s_i=\pm 1$).
The first term is the interaction term with a coupling constant $J_{ij}$, 
and the second term is the Zeeman energy with an applied magnetic field $h_i$.
In the case of a QAM~\cite{Dickson,Boixo1,Boixo2,Bermeister,IonTrap}, 
a tunneling term is added and expressed by
\begin{equation}
H=A(t) \sum_i \Delta_i  \sigma_i^x
+ B(t) [\sum_{i<j} J_{ij} \sigma_i^z \sigma_j^z 
+ \sum_i h_i \sigma_i^z ]
\label{QAM}
\end{equation}
where the variables are expressed by Pauli matrices $\sigma_i^\alpha$ ($\alpha=x,z$) instead of digital bits. 
The tunneling term is controlled such that it disappears at the end of the calculation.
Thus, $A=1$ and $B=0$ initially and $A=0$ and $B=1$ finally.
Although the Hamiltonian (\ref{QAM}) can be found in many physical systems in nature,
the tunneling term and the Ising term should be 
controlled separately and locally by electric gates to realize QAMs.

The advantage of superconducting qubits lies in the long coherence time 
of superconducting states~\cite{Dwave}.
Compared with the advance of superconducting qubits, 
the development of semiconductor qubits seems slower~\cite{Ladd}.
In semiconductor systems, spin and charge degrees of freedom 
can provide the qubit mechanism.
A qubit using spin is called a spin qubit and that using charge is called a charge qubit~\cite{book}. 
Spin qubits have longer coherence time because their independence from 
the noisy environment is greater than that of charge qubits.
For a realistic machine, it is crucially important to have a sufficient number of qubits.
Although two spin qubits are sufficiently controlled~\cite{Veldhorst,Maune,Kawakami}, 
the qubit operations more than three and more qubits have not yet been well succeeded
because the difficulty of controlling qubit-qubit interactions. 
The greatest advantage of using semiconductor devices is the possibility that 
the smallest artificial structures at the highest density can be manufactured in factories.
However, the spin qubits have not yet relished this benefit.
We think that charge qubits~\cite{Hayashi,Shinkai,Gorman,Ward,NEC,Valiev,Brandes,Fujisawa,Petta,Zhang,Shi,Mark} are better than spin qubits from the viewpoint 
of an integration of qubits, because charge qubits are interacting 
with mutual Coulomb interaction through capacitive couplings.
Thus, although charge qubits have generally less coherence time than spin qubits,
here we rather consider a charge qubit system 
by reviewing the proposal of a QAM using the conventional NAND flash memory
consisting of  floating gate (FG) cells in Ref.~\cite{tanamotoQAM}, 
and we extend the structure of a qubit to a coupled quantum dot (QD) system.
Coherent control of charge qubits using semiconductor QDs has been  
demonstrated in Ref.~\cite{Hayashi}, 
and coherent dynamics of two qubits based on coupled quantum dots (CQDs) has
enabled two-qubit operations in Ref.~\cite{Shinkai}.
Silicon charge-qubit operation has been experimentally 
shown in Ref.~\cite{Gorman} using a single-electron effect. 
New types of coupling in charge qubits have been experimentally investigated 
in Ref.~\cite{Ward}.
The smaller the semiconductor devices becomes, the larger the quantized energy intervals 
are expected to become, which is expected to result in increased coherence.
The fact that the fabrication technologies for semiconductor devices continue to progress 
is also beneficial to charge qubits.

Now, the cell size of advanced 2D NAND flash memory~\cite{Masuoka,Samsung,Toshiba} is less than 15 nm~\cite{iedm2012,iedm2013}, 
and the transistor size are entering into the quantum region below 7 nm\cite{TSMC7nm,Samsung7nm,IBM7nm}. 
The disadvantage of the charge qubit's short coherence time is 
expected to be reduced as transistor size decreases.
There are two important points that 2D NAND flash memory can be used 
as a good candidate of the charge qubit system.
First, in flash memory with 15 nm cells, single-electron effects 
can be observed at room temperature~\cite{Nicosia}.
The second point is that the inevitable interence effects 
between FG cells can be used as the interaction between qubits.
In commercial 2D NAND flash memories~\cite{Toshiba,Samsung,iedm2013,iedm2012} 
the distance between FG cells is of the same order as the size of the FG cells.
Thus, interference effects between FG cells are a major issue 
in present NAND flash memories~\cite{Lee}.  
In order to reduce the interference 
between FG cells, air-gap technologies are used~\cite{iedm2013} 
because the dielectric constant of air is smaller than that of tunneling oxide 
such as SiO${}_2$ (which has a dielectric constant of 3.8). 

NAND flash memories have a dominant share of the growing market for storage applications
extending from mobile phones to data storage devices in data centers~\cite{Takeuchi}. 
The NAND flash memories have the advantages of high-density memory capacity 
and low production cost per bit with low power consumption and high-speed programming and erasing mechanisms.
Now data storage of personal computers is also transitioning from hard disk to flash memory.
An FG cell corresponds to 1 bit for a single-level cell and $m$ bits for a multi-level cell. 
Each FG is typically made of highly doped polysilicon and placed in the middle of a gate insulator of a transistor~\cite{Brown,Aritome}. 
If there is no extra charge in an FG, the cell behaves like a normal transistor. 
In the programming or writing step, electrons are injected into the FG by 
applying voltage to the control gate. In the erasing step, electrons are 
ejected from the FG to the substrate by applying voltage to the back gate.
The amount of the charges of the FG determines the threshold voltage above which the 
current between the source and drain changes. 
In NAND flash memories, the FG cells are connected like a NAND gate circuit.
In general, the distance between the FGs is of the same order as 
the size of the FG, realizing a high-density memory.
For example, Sako {\it et al}. developed 64 Gbit NAND flash memory
in 15 nm CMOS technology~\cite{Toshiba}, which 
is organized by a unit of 16 KB bit-lines $\times$ 128 word-lines.
This means that the number of closely arrayed FG cells in a single unit is 16KB $\times$ 128 $\approx$ 2 MB.
The integration and miniaturization of flash memory cells have progressed continuously and 
the current flash memories have stacked 3D structures using trapping layers~\cite{Samsung2,Toshiba2}.

First, we theoretically show that a two-dimensional (2D)  FG array can be used as a QAM.
The QAM proposed here has the structure shown in Fig.1.
The FG cells are capacitively connected to each other, which is the same arrangement as that in a commercial flash memory.
The fundamental idea is that we will be able to regard a small FG cell in the single-electron region as a charge qubit. 
The size of the current FG NAND flash memory is 15 nm~\cite{iedm2012,iedm2013}, 
but it can be shrunk to less than 7 nm\cite{TSMC7nm,Samsung7nm,IBM7nm}. 
When the doping concentration of electrons is $5\times10^{18}$ cm${}^{-3}$, 
the number of electrons in a volume of 10$\times$10$\times$30 nm${}^{3}$
is about 15 and countable.
Once we can control the single-electron effects, we can realize a two-level 
system by using a crossover region between two different quantum states 
with different numbers of electrons, following, for example, Ref.~\cite{Makhlin}.
\begin{figure}
\centering
\includegraphics[width=8.0cm]{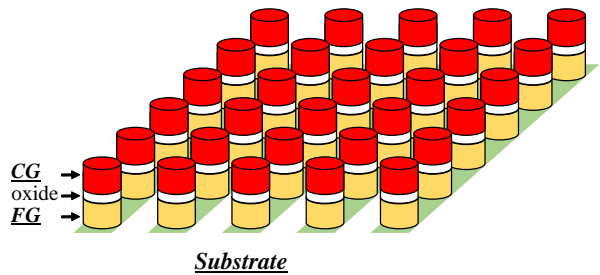}
\caption{
A quantum annealing machine (QAM) based on a floating gate (FG) array 
has the same structure as commercial NAND flash memory other than 
the thickness of tunneling barrier.
Control gates (CGs) are connected to bit-lines and word-lines 
arranged over the cell array.
A 2D FG memory array is constructed by arranging many blocks on the substrate.   
The interference effect between the FG cells is the origin of the Ising interaction.
Because we can use the same process as for conventional memory, 
the production cost is expected be low 
and the fabrication process is well established.
}
\label{fig1}
\end{figure}
Even when we can use state-of-the-art fabrication technologies, 
it is still difficult to control charge qubits with perfect coherence.
In general quantum computations, accurate control of wave functions is required 
from initial states to final states for measurements.
On the contrary, in a QAM, 
the  condition of the strict control of wave functions can be loosened 
provided that the final state is an eigenfunction of the target Hamiltonian, 
and the intermediate processes can include disturbance with several kinds of noise.
Thus, in the application of a QAM, there will be the advantages of small semiconductor devices 
such as high integration and productivity.

In Ref~\cite{tanamotoQAM}, we have shown that the FG array in 2D NAND flash memories 
can constitute a QAM by using the capacitive coupling between neighboring cells.
In this setup, the physical interactions are limited to the neighboring qubits.
On the other hand, in order to solve general combinatorial problems, 
the connection of all qubits to any qubits 
(all-to-all connection) should be prepared.
There are two major methods of realizing all-to-all connection based 
on solid-state qubits that have interactions only between neighboring qubits.
The Minor Embedding (ME) method by Choi~\cite{ME1,ME2} is used for the "Chimera" graph structure 
in the D-wave machine where logical qubits are replaced by chains of physical qubits.
Lechner, Hauke, and Zoller (LHZ)~\cite{LHZ} proposed
an alternative embedding method in which pairs of logical spins correspond to physical spins. 
T. Albash {\it et al} reported that the ME method showed better performance~\cite{Albash2}, 
although the possibility remains that the LHZ method will be improved in the 
future.
In this paper, we mainly consider how to implement the ME method 
in a 2D charge qubit array.

The remainder of this paper is organized as follows:
In Sec.~\ref{sec:formalism}, we briefly review the derivation of the Ising Hamiltonian 
from an FG array.
In Sec.~\ref{sec:CQD}, we explain how a QAM is implemented using CQD array.
In Sec.~\ref{sec:all2all}, we discuss how to realize all-to-all connection in the semiconductor qubits.
In Sec.~\ref{sec:decoherence}, we briefly discuss the effects of noise and decoherence on our QAM.
We close with a summary and conclusions in
Sec.~\ref{sec:conclusion}. 

\begin{figure*}
\centering
\includegraphics[width=14.0cm]{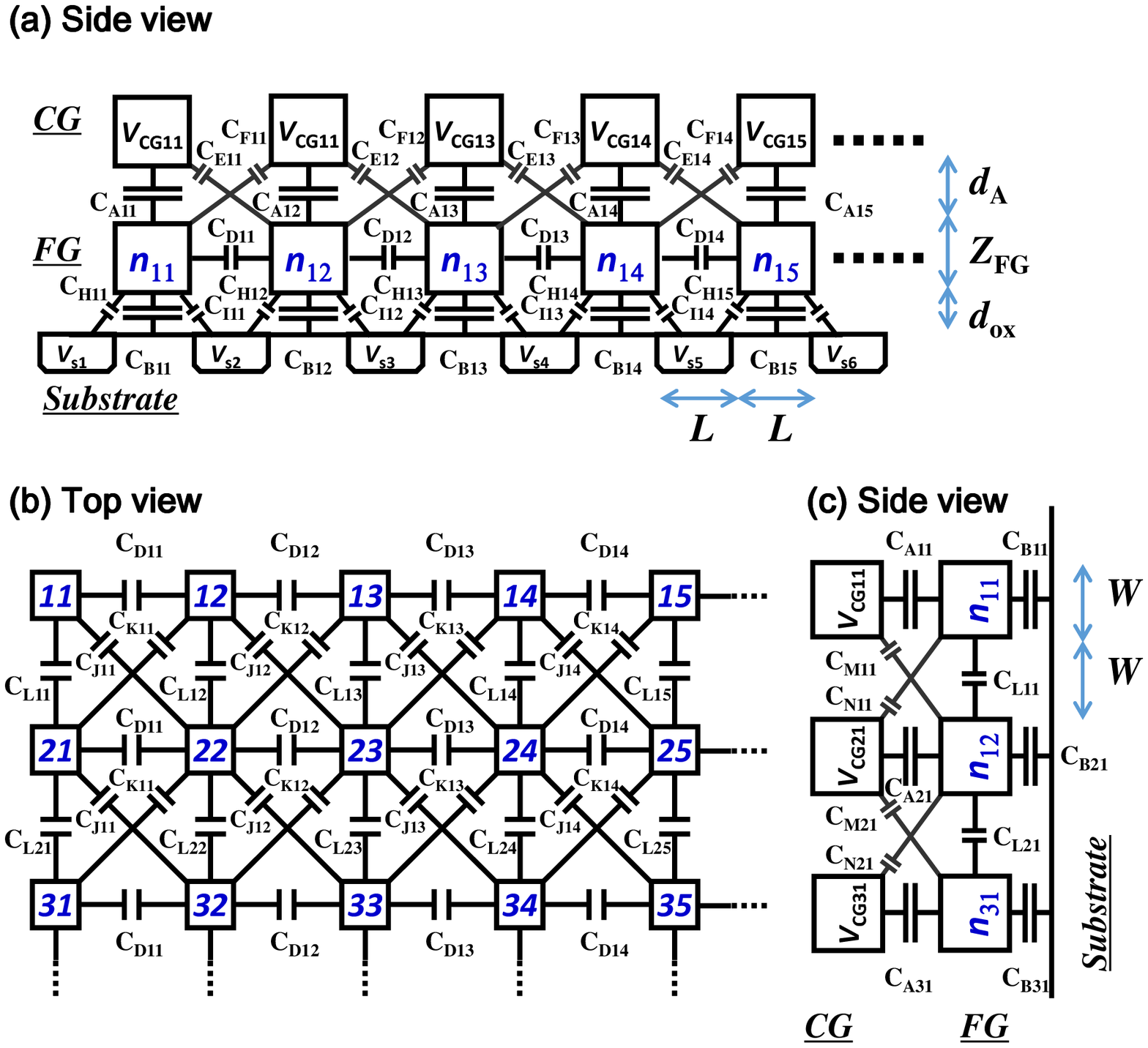}
\caption{
Schematics of the capacitance network model of 
2D FG array. We assume that the number of 
electrons in the floating gate (FG) is countable
and denoted by $n_{ij}$, which is controlled 
by the applied voltages on the control gate (CG) and 
substrate.
The size of FGs and the distance between FGs are
the same value of $L=W$.
(a) Side view of FG array.
(b) Top view of FG array. 
(c) Side view of FG array.
}
\label{fig2}
\end{figure*}
%
\section{General formalism of Ising interactions}\label{sec:formalism}
\begin{figure}[h]
\centering
\includegraphics[width=6.5cm]{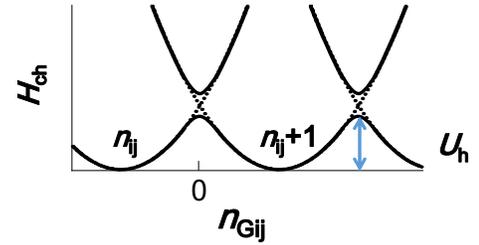}
\caption{
Schematic of the charging energy of a single FG as a function of $n_{G_{ij}}$, which 
represents the gate voltage of the $(ij)$-th cell~\cite{Makhlin}.
The crossover region $n_{G_{ij}}\approx 0$ of two parabolas
realizes a two-state quantum system.
$U_h$ indicates the typical scale at which single-electron effects can be controlled.
}
\label{fig3}
\end{figure}

\begin{figure}[h]
\centering
\includegraphics[width=8.5cm]{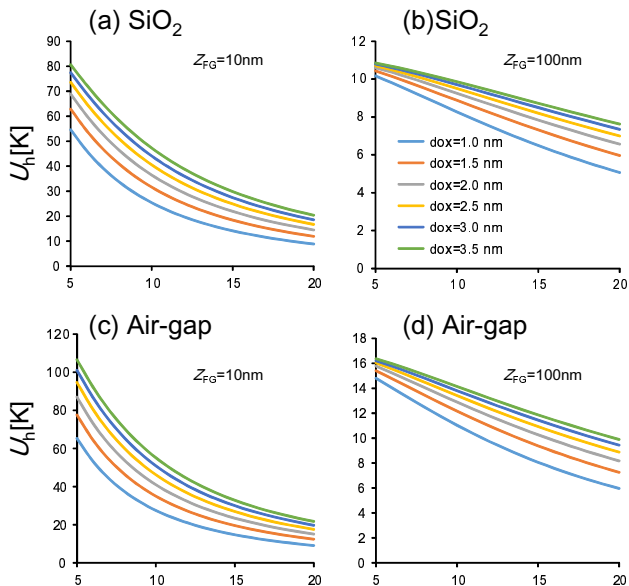}
\caption{
$U_h$ that determines the single-electron effect
(Fig.~\ref{fig3}) is calculated as a function of the size of the FG.
(a)(c) $U_h$ for the FG height of $Z_{\rm FG}$=10 nm. (b)(d) $U_h$ for $Z_{\rm FG}$=100 nm.
In (a) and (b), all insulating materials are assumed to be SiO${}_2$.
In (c) and (d), we assume that there are no materials in the centers of four qubits (air-gap structure). 
The thickness of the
control gate oxide depends on $d_{\rm ox}$ and the coupling ratio is
CR = 0.3.}
\label{fig4}
\end{figure}

Here, we derive an Ising Hamiltonian from the coupled 2D FG array by using a capacitance network model~\cite{Pavan} 
as shown in Fig.~\ref{fig2}. 
Each cell consists of an FG and a control gate. 
We assume a Coulomb blockade regime 
and the number of electrons of the FG cell at position $(i,j)$ is expressed by $n_{ij}$.
We can also define charge  states $|n_{ij}\ra$.
FGs are capacitively  connected to the nearest FG, CG, and substrate with source and drain.
The Ising interaction comes from charging energy. 
The charging energy of the 2D system of Fig.~\ref{fig2} is given by 
\begin{eqnarray*}
U\!\!&=&\!\!\sum_{ij} \biggl\{
 \fc{q_{A_{ij}}^2}{2C_{A_{ij}}} 
+\fc{q_{B_{ij}}^2}{2C_{B_{ij}}} 
+\fc{q_{H_{ij}}^2}{2C_{H_{ij}}} 
+\fc{q_{I_{ij}}^2}{2C_{I_{ij}}} 
\nonumber  \\
&+&\left[
 \fc{q_{D_{ij}}^2}{2C_{D_{ij}}}
+\fc{q_{E_{ij}}^2}{2C_{E_{ij}}} 
+\fc{q_{F_{ij}}^2}{2C_{F_{ij}}} \right] 
\nonumber \\
&+&\left[
 \fc{q_{L_{ij}}^2}{2C_{L_{ij}}} 
+\fc{q_{M_{ij}}^2}{2C_{M_{ij}}}  
+\fc{q_{N_{ij}}^2}{2C_{N_{ij}}} \right]  
+\fc{q_{J_{ij}}^2}{2C_{J_{ij}}} 
+\fc{q_{K_{ij}}^2}{2C_{K_{ij}}}  
\nonumber \\
&-&(q_{A_{ij}}+q_{E_{ij}}+q_{M_{ij}}
+[q_{F_{ij}}+q_{N_{i-1,j}}])V_{{\rm CG}_{ij}}
\nonumber \\
&+&q_{B_{ij}}V_{{\rm sub}_{ij}}
+q_{H_{ij}}V_{s_{ij}}
+q_{I_{ij}}V_{d_{ij}}
\nonumber \\
&+&\bigl( 
 -q_{A_{ij}}+q_{B_{ij}}
 +q_{H_{ij}}+q_{I_{ij}} 
\nonumber \\
&+&[q_{D_{ij}}-q_{D_{i,j-1}}
 -q_{F_{ij}}-q_{E_{i,j-1}}]
\nonumber \\
&+& 
 [q_{L_{ij}}-q_{L_{i-1,j}}-q_{N_{ij}}-q_{M_{i-1,j}}]
\nonumber \\
 &+&q_{J_{ij}}-q_{J_{i-1,j-1}} 
 +q_{K_{i,j-1}}-q_{K_{i-1,j}} 
 -N_{ij} \bigr)\lambda_{ij}  
 \biggr\},
\end{eqnarray*}
where $q_{A_{ij}}$,...,$q_{N_{ij}}$ are stored charges on capacitances.
The charge distribution is obtained after minimizing  
$U$ by adjusting the Lagrange multipliers $\lambda_{ij}$.
When we neglect the interaction beyond the next-neighboring interactions, 
after the simple but long calculations, the charge distribution is obtained and given by
\begin{equation}
U \approx \sum_{ij} \left[ \frac{Q_{ij}'^2}{2} -W_{ij} \right],
\label{app2_U}
\end{equation}
where 
\begin{eqnarray}
Q'_{i,j}  \!&\equiv& \!\{Q_{i,j}
+Q'_{i,j-1}C'_{D_{i,j-1}} 
+Q'_{i-1,j-1}C'_{J_{i-1,j-1}}
\nonumber \\
&+&Q'_{i-1,j}C'_{L_{i-1,j}}
+Q'_{i-1,j+1}C'_{K_{i-1,j}} \}/\sqrt{C'_{a_{i,j}}},
\label{app2_q}\\
Q_{i,j}\!&\equiv &\!
Q^0_{v_{ij}}+N_{ij},
\label{app2_q2}\\
C'_{a_{i,j}} \!\!&\!\!\equiv\!\!&\!\!  
C_{a_{i,j}} -\{C'_{J_{i-1,j-1}}\}^2 -\{C'_{L_{i-1,j}}\}^2
\nonumber \\
&-&\{C'_{D_{i,j-1}}\}^2  
-\{C'_{K_{i-1,j}}\}^2,  
\\
C_{a_{i,j}}\!&\equiv &\!
  C_{A_{ij}} +C_{B_{ij}}+  C_{H_{ij}}+  C_{I_{ij}}  
+[C_{D_{ij}}+C_{F_{ij}}]
\nonumber \\
\!&+&\![C_{L_{ij}}+C_{N_{ij}}]
+ C_{J_{ij}}
\nonumber \\
\!&+&\! C_{D_{i,j-1}}+C_{E_{i,j-1}}+ C_{K_{i,j-1}}
\nonumber \\
\!&+&\!C_{K_{i-1,j}}+C_{L_{i-1,j}}+C_{M_{i-1,j}} 
+C_{J_{i-1,j-1}},
\\
C'_{D_{i,j}}  \!&\equiv&\! \{
C_{D_{i,j}}
+C'_{L_{i-1,j}}C'_{J_{i-1,j}} 
\nonumber \\
&+&C'_{L_{i-1,j+1}}C'_{K_{i-1,j}}
\}/\sqrt{C'_{a_{i,j}}},
\\
C'_{L_{i,j}} \!&\equiv&\!  \{
C_{L_{i,j}}            
 +C'_{D_{i,j-1}}C'_{J_{i,j-1}}
 \}/\sqrt{C'_{a_{i,j}}},
\\
C'_{J_{i,j}}\!&\equiv&\!  \{
C_{J_{i,j}} \}/\sqrt{C'_{a_{i,j}}},
\\
C'_{K_{i,j}} \!&\equiv&\! \{
C_{K_{i,j}}            
 +C'_{D_{i,j}}C'_{L_{i,j}}
 \}/\sqrt{C'_{a_{i,j+1}}},
\\
W_{i,j} \!&\equiv &\! 
 C_{A_{ij}}V_{{\rm CG}_{ij}}^2
+C_{B_{ij}}V_{{\rm sub}_{ij}}^2
+C_{H_{ij}}V_{s_{ij}}^2
+C_{I_{ij}}V_{d_{ij}}^2 
\nonumber \\
\!&+&\!(C_{E_{ij}}+C_{M_{ij}}
+C_{F_{i,j-1}}
+C_{N_{i-1,j}}
)V_{{\rm CG}_{ij}}^2,
\end{eqnarray}
with 
\begin{eqnarray}
Q^0_{v_{ij}}&\equiv& 
 C_{A_{ij}}V_{{\rm CG}_{ij}}
+C_{B_{ij}}V_{{\rm sub}_{ij}}
+C_{H_{ij}}V_{s_{ij}}
+C_{I_{ij}}V_{d_{ij}} \nonumber \\
&+& 
C_{F_{ij}}V_{{\rm CG}_{i,j+1}}
+C_{N_{ij}}V_{{\rm CG}_{i+1,j}}  
+C_{E_{i,j-1}}V_{{\rm CG}_{i,j-1}}
\nonumber \\
&+&C_{M_{i-1,j}}V_{{\rm CG}_{i-1,j}}.
\end{eqnarray}
Here, we have continuously completed the squares to find the minimum points of the 
charging energy. 
Concretely, starting from $Q_{11}'\equiv Q_{11}/\sqrt{C_{a_{11}}}$, we take
$Q_{12}'\sqrt{C_{a_{12}}} \equiv Q_{12}+Q_{11}'C_{D_{1,1}}'$ and so on.
Note that $Q_{ij}'$s in Eq.~(\ref{app2_q}) include
$Q'_{i,j-1}$, 
$Q'_{i-1,j-1}$, 
$Q'_{i-1,j}$ and
$Q'_{i-1,j+1}$. Thus, the form of Eq.~(\ref{app2_U}) generates 
the Ising interactions between neighboring FG cells through Eq.~(\ref{app2_q2}).
And thus, the Ising interactions and Zeeman terms are 
obtained as a result of the parabolic form of the charging energy.
Following Ref~.\cite{Makhlin}, 
the superposition state is constructed around the region
\begin{equation}
(n_{ij}+Q_{v_{ij}}^0)^2 = (n_{ij}+1+Q_{v_{ij}}^0)^2.
\label{parabola_cond}
\end{equation}
This is the region where the charging energy of $n_{ij}$ electrons 
equals  that of the $n_{ij}+1$ electrons, 
and quantum states $|n_{ij}\ra$ and $|n_{ij}+1\ra$ states can be defined 
as shown in Fig.~\ref{fig3}. 
We use the effective gate voltage $n_{G_{ij}}$ given by
\begin{equation}
n_{ij}+Q_{v_{ij}}^0= n_{G_{ij}} -1/2.
\end{equation}
($n_{G_{ij}} \ll 1$).
For this region, we can approximate the following equation
\begin{eqnarray}
\sum_{m=0}^{1}(n_{ij}+m+Q_{v_{ij}}^0)^2 & \!\rightarrow \! &\!
\frac{1}{2}  n_{G_{ij}} \sigma_{ij}^z +\left(n_{G_{ij}}^2 +\frac{1}{4}\right) I_{ij}, \ \ \ \ \ 
\end{eqnarray}
where $\sigma^z$, $I_{ij}$ are Pauli matrix and unit matrix, respectively,  
based on the $\{|n_{i'j'} \ra,|n_{ij} \ra \}$ system.
Thus, the charging energy term as a function of $n_{G_i}$ transforms to
\begin{equation}
U=\sum_{i,j} h_{ij} \sigma_{ij}^z 
+\sum_{i<j,i'<j'} J_{\{ij\}\{i'j'\}} \sigma_{ij}^z \sigma_{i'j'}^z  +{\rm Const}
\label{U},
\end{equation}
where the Ising interactions are given by
\begin{eqnarray}
\!J_{\{ij\}\{i,j+1\}}\!&\!\!\approx\!\!&\!\frac{C_{D_{ij}}}{4C'_{a_{ij}}C'_{a_{i,j+1}}}, 
\\
\!J_{\{ij\}\{i+1,j\}}\!&\!\!\approx\!\!&\!\frac{C_{L_{ij}}}{4C'_{a_{ij}}C'_{a_{i+1,j}}}, 
\\
\!J_{\{ij\}\{i+1,j+1\}}\!&\!\!\approx\!\!&\!\frac{C_{J_{ij}}}{4C'_{a_{ij}}C'_{a_{i+1,j+1}}}, 
\\
\!J_{\{i+1,j\}\{i,j+1\}}\!&\!\!\approx\!\!&\!\frac{C_{K_{i,j}}}{4C'_{a_{i+1,j}}C'_{a_{i,j+1}}}, 
\end{eqnarray}
The magnetic field is given by
\begin{eqnarray}
\!h_{ij} \!&=\!& \! \frac{1}{2C'_{a_{ij}}} \!
\Biggl[1\!
+\!\frac{C_{D_{ij}}^2}{C'_{a_{ij}}C'_{a_{i,j+\!1}}}
+\!\frac{C_{L_{ij}}^2}{C'_{a_{ij}}C'_{a_{i+\!1,j}}}
+\!\frac{C_{J_{ij}}^2}{C'_{a_{ij}}C'_{a_{i+\!1,j+\!1}}} 
\nonumber \\
&+&
\!\frac{C_{K_{i,j-1}}^2}{C'_{a_{ij}}C'_{a_{i+\!1,j-\!1}}} 
\Biggr]  n_{G_i} 
\nonumber \\
&+&\frac{ C_{D_{i,j-1}} n_{G_{i,j-1}}}{2C'_{a_{i,j-1}}C'_{a_{i,j}}}
 + \frac{ C_{D_{ij}} n_{G_{i,j+1}} }{2C'_{a_{i,j}}C'_{a_{i,j+1}}} 
 \nonumber \\
&+&\frac{ C_{J_{i-1,j-1}} n_{G_{i-1,j-1}}}{2C'_{a_{i-1,j-1}}C'_{a_{i,j}}}
 + \frac{ C_{J_{ij}} n_{G_{i+1,j+1}} }{2C'_{a_{i,j}}C'_{a_{i+1,j+1}}} 
\nonumber \\
&+&\frac{ C_{L_{i-1,j}} n_{G_{i-1,j}}}{2C'_{a_{i-1,j}}C'_{a_{i,j}}}
 + \frac{ C_{L_{ij}} n_{G_{i+1,j}} }{2C'_{a_{i,j}}C'_{a_{i+1,j}}} 
\nonumber \\
&+&\frac{ C_{K_{i-1,j}} n_{G_{i-1,j+1}}}{2C'_{a_{i-1,j+1}}C'_{a_{i,j}}}
 + \frac{ C_{K_{i,j-1}} n_{G_{i+1,j-1}} }{2C'_{a_{i,j}}C'_{a_{i+1,j-1}}},
\label{ngh} 
\end{eqnarray}
and
\begin{eqnarray}
\sigma_{ij}^x&=& |n_{ij} \ra \la n_{ij}+1 | + |n_{ij}+1 \ra \la n_{ij}|,
\\
\sigma_{ij}^z&=&-|n_{ij} \ra \la n_{ij} | + |n_{ij}+1 \ra \la n_{ij} +1|,
\\
I_{ij}&=&|n_{ij} \ra \la n_{ij} | +|n_{ij}+1 \ra \la n_{ij} +1|.
\end{eqnarray}
Thus, the 2D FG array can be mapped to a 2D Ising spin system 
with antiferromagnetic couplings.
In this 2D case, there are next-nearest couplings by $C_{J_{ij}}$ and $C_{K_{ij}}$.
These next-nearest couplings induce spin states conflict with 
those induced by nearest neighboring couplings. 
We can erase the next-nearest couplings by inserting 
the air-gap~\cite{iedm2013} in the middle of four FG cells.
This is because the dielectric constant of air is lower than that 
of SiO${}_2$, we can reduce the effect of the capacitance couplings.
In Ref.~\cite{tanamotoQAM}, numerical estimations 
based on 1D capacitance network model were carried out.
Those calculations showed that the smaller size of FGs 
induces higher temperature operations as expected.
In Ref.~\cite{tanamotoQAM}, 
Technology CAD(TCAD) tools were also used and 
it was shown that the response speed of the FGs are 
in order of 10${}^{-11}$ s for $L=W=15$ nm FGs.
Here, let us check the effect of the next-nearest coupling 
by calculating the $U_h$, 
which corresponds to the height of the charging energy $U_h$
and is calculated by the coefficient of $n_{G_i}$ in Eq.(\ref{ngh}) such as
\begin{eqnarray}
U_h &\equiv & 
\frac{1}{8C_{a_{ij}}} \!
\Biggl[1\!
+\!\frac{C_{D_{ij}}^2}{C_{a_{ij}}C_{a_{i,j+\!1}}}
+\!\frac{C_{L_{ij}}^2}{C_{a_{ij}}C_{a_{i+\!1,j}}}
+\!\frac{C_{J_{ij}}^2}{C_{a_{ij}}C_{a_{i+\!1,j+\!1}}} 
\nonumber \\
&+&
\!\frac{C_{K_{i,j-1}}^2}{C_{a_{ij}}C_{a_{i+\!1,j-\!1}}} 
\Biggr]
\end{eqnarray}
Figure \ref{fig4} shows an example of the numerical calculations of 
$U_h$ surrounded by qubits
when the coupling ratio $CR$ is given by 0.3.
The coupling ratio 
indicates the degree of controllability of the gate electrode.
and given by 
\begin{equation}
CR=\frac{C_A}{C_A+C_B},
\label{CR}
\end{equation}
as it is frequently used in the field of FG memory. 
We consider a FG whose size $L$ and width $W$ have the same value, $L=W$. 
When the thickness of the tunneling oxide is $d_{\rm ox}$, 
the thickness of the insulator between the FG and the CG is 
given by
\begin{equation}
d_{\rm A}=d_{\rm ox}(1-CR)/CR. 
\end{equation}
The capacitances are defined by using their simplest expressions given by
\begin{eqnarray}
C_A&=& \epsilon_{\rm sio2} \epsilon_0 LW/d_{\rm A}, \\ 
C_B&=& \epsilon_{\rm sio2} \epsilon_0 LW/d_{\rm ox}, \\
C_D&=& \epsilon_{\rm sio2} \epsilon_0 Z_{\rm FG}W/L;  \\
&...& \nonumber \\ 
C_J &=&C_K=C_D/ \sqrt{2},
\end{eqnarray}
where $\epsilon_{\rm sin}=3.9$ and $\epsilon_0=8.854 \time 10^{-12}$F/m.
For the case of the air-gap in the middle of the four qubits,
we use
\begin{eqnarray}
C_J &=& C_K=\epsilon_0 Z_{\rm FG}W/(L\sqrt{2}).
\end{eqnarray}
Figure \ref{fig4} shows that weak next-nearest interactions,
which corresponds to the air-gap cases ((c)(d)), 
enhance the single electron effects, compared with the 
existence of full next-nearest interactions of the SiO${}_2$ cases ((a)(b)).
From Fig.~\ref{fig4}, 3\%-30\% increase in $U_h$ can be seen
for $Z_{\rm FG}=10$nm and 20\%-50\% increase in $U_h$ can be seen 
for $Z_{\rm FG}=100$nm. 

As long as the capacitance network model is used, $J$, $U_h$ and other physical parameters depend on only capacitances. 
In semiconductor systems, however, the capacitance changes depending on applied voltages. 
As an example, the capacitance of PN junction changes depending on the 
change of the depletion region as the applied voltage changes~\cite{Grove}.
In the present structure, the NAND flash memory includes many different regions 
of n-type and p-type semiconductors. 
Thus, it is possible that the capacitances change complicatedly when the applied bias is changed.
The detailed dependence of the change of capacitances 
will be estimated by carrying out TCAD simulations.
This will require a lot of calculations and considerations, and be future issues.

\subsection{Analytical formation of tunneling term}
The tunneling term Eq.~(\ref{QAM})
is derived by using the Wentzel-Kramers-Brillouin (WKB) approximation~\cite{Harrison}, 
and given by
\begin{eqnarray}
\! \!
\lefteqn{ \sum_{k,k'} \Delta_{k,k'} c_{k_R}^\dagger c_{k_L} 
=N_LN_R  \frac{m_0R_y}{ m_{\rm si}^*} 
\left[\frac{\pi a_0 }{L}\right]^2  }
\nonumber \\
&\times &
\exp \left\{\left[ -\frac{d_{\rm ox}}{a_0} 
\sqrt{\frac{m_{\rm ox}^*(V_{\rm ox}-E_{\rm F}')}{m_0R_y} } \right]\right\}
c_{k^R_F}^\dagger c_{k_F^L},
\label{Tunnel}
\end{eqnarray}
where $V_{\rm ox}\approx$ 3.0 eV is the potential height of the tunneling barrier, 
$m_{\rm si}^*\approx 0.19m_0$ and $m_{\rm ox}^*\approx 0.5m_0$ are the effective masses of electrons in Si and tunneling barrier, respectively
($m_0$is an electron mass in vacuum).
$c_{k_L}$ and  c$_{k_R}$ are annihilation operators of both sides of the tunneling barrier.
$k_F$ is the wave vector at the Fermi energy. $a_0\approx 0.0529$ nm is the Bohr radius and 
$R_y\approx 13.6$ eV is the Rydberg constant. 
$N_L$ and $N_R$ are the numbers of electrons on the two sides of the tunneling barrier 
that participate in the tunneling event.
This tunneling term is a function of the gate voltage $V_{\rm CG}$ 
depending on the 
shift in Fermi energy $E_F$, such as $E_{\rm F}'=E_{\rm F}+V_{\rm CG}$,
and $E_F$ is calculated from the FG doping concentration.
When $E_F'$ increases, the effective tunneling barrier is lowered and the tunneling rate increases (switches on). 
Conversely, when $E_F'$ decreases, the effective tunneling barrier is raised 
and the tunneling switches off.
Thus, the tunneling can be switched on or off by controlling the gate and substrate bias.
Eq.~(\ref{Tunnel}) is the expression of the tunneling from the approach of many electrons.
The smaller the number of electrons in FGs becomes, the better coherence of qubits is expected.
Thus, in such case of a smaller number of electrons, 
we will have to construct more elaborate formulation in the future.

\section{QAM based on CQD}\label{sec:CQD}
\begin{figure}[h]
\centering
\includegraphics[width=8.5cm]{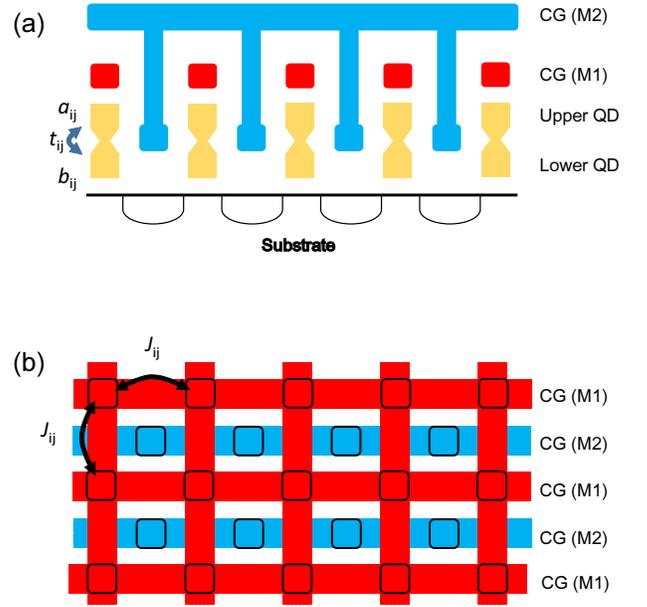}
\caption{
(a) Side view of CQD array. $\hat{a}_{ij} (\hat{b}_{ij})$ describes the annihilation operator 
when an extra electron exists in the upper (lower) QD. $t_{ij}$ is the tunneling rate
between the two QDs.
(b) Top view of CQD array. 
The red line is the first wiring layer (M1)
that control the charge distribution of CQD. 
the blue line is the second wiring layer (M2)
that works as the split-gate to control the 
tunneling rate $t_{ij}$ between the upper QD and 
lower QD. 
}
\label{fig5}
\end{figure}

As mentioned in the introduction, 
because it is very difficult to uniformly construct small structures of nm scale,
a simpler structure is better for fabrication processes.
In this meaning, NAND flash memory is best for QAM from the viewpoint of its simple structure.
Here, we consider a QAM of a little bit more 
complicated structure based on coupled quantum dots (CQDs).
CQDs or double QDs have been 
widely investigated in the field of nano-physics~\cite{JPSJ1,JPSJ2,JPSJ3,JPSJ4,Nishiguchi,tana2}.
Depending on whether an extra electron exists in one QD or the other QD, 
the logical states $|0\ra$ and $|1\ra$ are defined.
Because we are focusing on the integration of qubits,  
we have to stack QDs as shown in Fig.~\ref{fig5}.
If we don't have to switch on/off the tunneling between two QDs, 
we can stack the simple oxide material such as SiO${}_2$ between two QDs. 
However, because quantum annealing process requires switching on/off of the 
tunneling between QDs, 
we need the structure that enables the switching on/off of the tunneling.
In the usual lateral QDs~\cite{JPSJ4}, 
two positions of the excess electron is usually connected 
by "split-gates".
The split-gates change the depth of the depletion layer and the tunnelings 
between QDs are controlled.
In the present case, we will have to embed the additional electrodes which 
work as the split-gate between CQDs as shown in Fig.~\ref{fig5}.
The difference from Ref.~\cite{tana0} is that there are electrodes between the CQDs.
Compared with a quantum computer,  
we do not have to switch tunneling on/off independently in the case of quantum annealer. 
Thus we can set a common electrode to the split-gates
to switch the tunneling on/off simultaneously.
For the CQDs, excess electrons are confined in the closed two QDs
and the two QDs are separated from electrodes, it is expected 
that the coherence time of CQD system becomes larger than that of QAM based on flash memory.
However, in the CQD system, we have to add additional split-gates, and the size 
between CQDs (qubits) becomes larger. 
These are considered to degrade the coherence. 
In the future, we will have to estimated these trade-off 
between QAMs based on flash memory and CQDs.  

\subsection{Hamiltonian of CQDs} 
Here we consider the Hamiltonian of the 2D arrayed CQDs by
starting from the tunneling Hamiltonian
\begin{equation}
H=\sum_{i,j=1}^N (t_{ij} \hat{a}^\dagger_{ij} \hat{b}_{ij}
+t_{ij}^*\hat{b}^\dagger_{ij} \hat{a}_{ij}
+\epsilon_{\alpha_{ij}}\hat{a}^\dagger_{ij} \hat{a}_{ij}
+\epsilon_{\beta_{ij}}\hat{b}^\dagger_{ij} \hat{b}_{ij})
+H_{ch},
\end{equation}
where $\hat{a}_{ij} (\hat{b}_{ij})$ describes the annihilation operator when the
excess electron exists in the upper (lower) QD, and 
$\epsilon_{\alpha_{ij}} (\epsilon_{\beta_{ij}})$
shows the electronic energy of the upper (lower) QD. 
$N$ is the number of CQDs.
Compared with the case of FG system Eq.(\ref{Tunnel}), 
we could describe the electronic system more microscopically.
$H_{\rm ch}$ is the charging energy 
of the CQD system. Because we consider the Coulomb blockade in
the weak coupling region, the operational temperature be less than the charging
energies. The operational speed should be less than the
CR constant of the capacitance network so that the double-well
potential profile generated by the charging energy is
effective adiabatic region. 
As for the interaction between qubits,
the distribution of the extra charge is considered to be
antiferromagnetic because of the repulsive Coulomb interaction.
Similarly to the FG case, we showed that this interaction between qubits is an Ising
interaction by minimizing a similar charging energy in Ref.~\cite{tana0,tana1}.
Thus, the Hamiltonian of the CQD system is given by
\begin{eqnarray}
H&=&\sum_{i,j=1}^N (t_{ij} \hat{a}^\dagger_{ij} \hat{b}_{ij}
+t_{ij}^*\hat{b}^\dagger_{ij} \hat{a}_{ij}
+\epsilon_{\alpha_{ij}}\hat{a}^\dagger_{ij} \hat{a}_{ij}
+\epsilon_{\beta_{ij}}\hat{b}^\dagger_{ij} \hat{b}_{ij})
\nonumber \\
&+&\sum_{i,j,i',j'} J_{ij,i'j'} \sigma_{ij}^z\sigma_{i'j'}^z.
\end{eqnarray}

\section{Toward all-to-all connection}\label{sec:all2all}
For connecting all qubits to any qubits, {\it i.e.}, all-to-all connection,
the two major methods are well known:
the Minor embedding method by Choi~\cite{ME1,ME2} and the method by LHZ~\cite{LHZ}.
Let us discuss how to realize these methods in the NAND flash memory system.

\subsection{Implementation of the minor embedding}
\begin{figure}
\centering
\includegraphics[width=8.6cm]{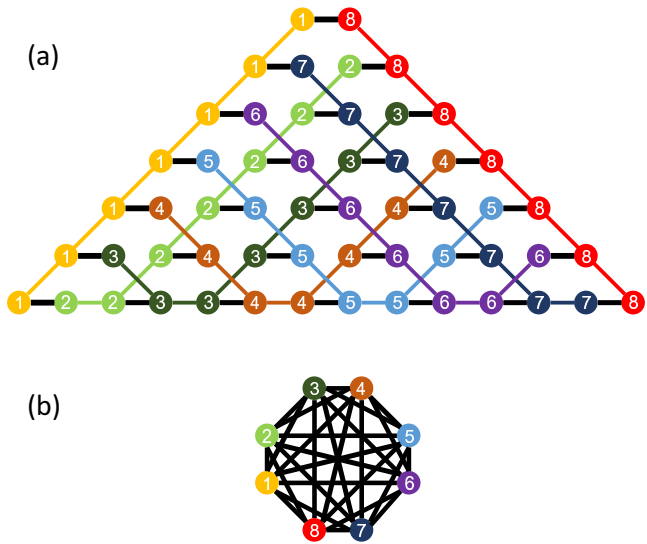}
\caption{
(a) Minor embedding method from Refs.\cite{ME1,ME2}.
(b) All-to-all connection realized by (a).
}
\label{fig6}
\end{figure}

\begin{figure}
\centering
\includegraphics[width=8.6cm]{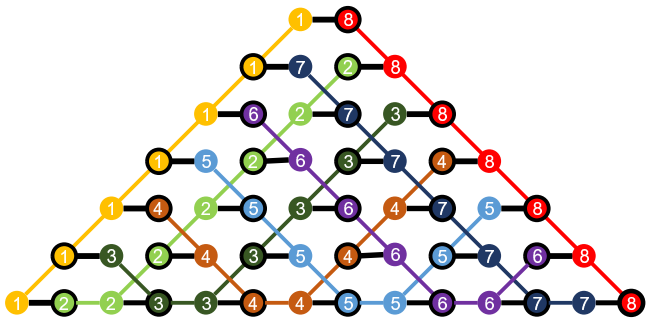}
\caption{
In the conventional MR method~\cite{ME1,ME2},
coupling between qubits in the same logical qubit is 
ferromagnetic. This figure shows the case when 
all interactions between qubits are antiferromagnetic
couplings. We can set data inversely.
Solid circles indicate the qubit in which data are input
inversely.
}
\label{fig7}
\end{figure}

Figure~\ref{fig6} shows the ME method of Ref.~\cite{ME1,ME2}.
Circles indicate qubits and solid lines indicate interactions between qubits.
In order to connect a spin with any distant spins, 
Choi~\cite{ME1,ME2} introduced a logical spin that consists of many spins with the same spin states.
In Fig.~\ref{fig6}, the qubits with the same number constitute a logical qubit.
The logical qubits are connected by strong interactions.
In Ref.~\cite{ME1}, the condition of the strength of the interaction 
in the qubits of a logical state $T_i$ ($i$-th tree) is expressed by
\begin{equation}
J >(|h_i|+\sum_{j \in {\rm nbr}(i)} |J_{ij}|).
\label{me_eq}
\end{equation}
This means that coupling $J$ in the logical spins ($T_i$) is stronger 
than the sum of the surrounding coupling plus the local magnetic field $h_i$.
Note that the interactions between logical qubits in Refs.~\cite{ME1,ME2} 
are ferromagnetic interactions.
On the other hand, the interactions described in the previous sections 
are antiferromagnetic interactions.
We consider that similar discussions as in Refs.~\cite{ME1,ME2} are possible. 
Because the ground state of a 1-D Ising antiferromagnetic chain is given by 
$|\uparrow\downarrow\uparrow\downarrow\uparrow....\ra$, 
we can apply this idea to the antiferromagnetic system as shown in Fig.~\ref{fig7}.
 
\begin{figure}
\centering
\includegraphics[width=3.5cm]{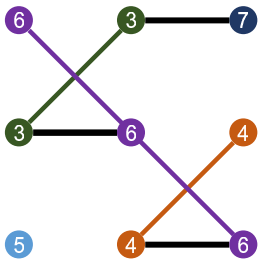}
\caption{
The typical part of the minor embedding method of 
Ref.~\cite{ME1,ME2}.
There are three types of bonding between two qubits.
The first type is a fixed coupling between qubits 
that have the same number in the figure.
This type of coupling is so strong that the qubits with 
the same number have the same spin state.
The second type is that there is no coupling between qubits.
The third type is that the coupling $J$ represents the data 
and should be changeable. In this figure, the qubits with different numbers
are connected by this third type. 
}
\label{fig8}
\end{figure}

Let us consider the implementation of the minor embedding(ME) method 
to the 2D Ising array with constant antiferromagnetic interactions 
using nano structures.
In the ME method, we have to prepare three types of 
bonding between two qubits as shown in Fig.~\ref{fig8}:
(i) The first type is a fixed coupling between qubits as shown in Eq.(\ref{me_eq}).
(ii) The second type is that there is no interaction between two qubits.
(iii) The third type is that the coupling $J$ represents the data 
and should be changeable.
Here, we consider possible forms of these types of bonding, 
focusing on the 2D qubit array with Coulomb interactions
through insulating materials. 
Then, the constant interaction of the first type is already realizable through 
the insulating material with relatively high dielectric constant 
such as SiO${}_2$. 
\begin{figure}
\centering
\includegraphics[width=8.6cm]{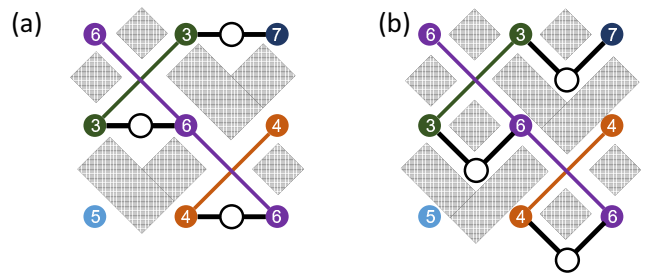}
\caption{
A possible form of the qubit array to realize the three types 
of bonding shown in Fig.~\ref{fig8}.
The first type of bonding between qubits are 
realized by natural Coulomb interactions between the cells.
The second type of bonding is realized by inserting 
low dielectric materials or air-gaps~\cite{iedm2013} to erase the Coulomb interactions
between cells. Shaded parts show the low dielectric materials or the air-gap. 
The third type is realized by setting control qubits.
The position of the control qubits are different between (a) and (b).
}
\label{fig9}
\end{figure}
The second type is realizable by replacing the 
materials between the qubits with ones that have low dielectric constants.
In the case of Si-based qubits, the air-gap~\cite{iedm2013} that means
the space between two qubits is filled with air is available to present interaction 
between the qubits. 
The third type can be realized by inserting a new 
qubit that controls the strength of the bonding
by using the method proposed in Ref.~\cite{Niskanen}.
It is considered that, by applying large oscillating bias, 
the magnitude of the interaction 
between the two qubits is controlled as shown in Ref.~\cite{Niskanen}.
Figures~\ref{fig9} (a) and (b) illustrate these proposals.
The shaded parts indicate the air-gap region 
where space is filled with air or the materials with low dielectric constants.
In Fig~\ref{fig9}(b), the intermediate qubits in the 
controlling bonds are placed in the same lattice structure 
with other qubits.  
This structure is applicable to the CQD system mentioned in Sec.2, 
because the split-gate electrodes are inserted in the middle of the four qubits.


\subsection{Application of LHZ method}
The LHZ method also enables connection of all qubits to any qubits~\cite{LHZ}.
In the LHZ method, $N$ logical Ising spins are encoded in
$M=N(N-1)/2$ physical qubits with constraints.
Each physical qubit represents the relative configuration of two logical spins such 
that the physical qubit takes the value 1 if the two connected logical
spins point in the same direction and 0 otherwise.
New constraints are introduced as four-body interactions 
or three body interactions 
to keep the consistency of qubit configurations.
Because the four-body interaction and three-body interaction are 
unnatural interactions, we have to generate this higher order 
interactions starting from natural two-body interactions.
Lechner showed that the four-body interaction can be realized 
by using a series of CNOT gates~\cite{Lechner}.
In order to realize the series of CNOT gates, sufficient quantum coherence
will be required.
Thus this method requires perfect control of the electronic system.

\section{Decoherence}\label{sec:decoherence}
In Ref.~\cite{tana0} and Ref.~\cite{tanamotoQAM}, we roughly estimated the coherence time of 
the charge qubit based on the spin-boson model~\cite{Legget}.
In Ref.~\cite{tana0} and Ref.~\cite{tanamotoQAM}, we considered
a low-temperature region, where only acoustic phonons
play a major role in the decoherence mechanism.
The interaction term between a qubit and acoustic phonons
is derived from that of amorphous SiO${}_2$~\cite{Garcia,Wurger}. 
The estimated coherences are given by 
around 4.8$\times 10^{-7}$ s, during which more than thousands of quantum
calculations can be realized if the switching time is less than nano seconds.
However, these estimations were carried out on the assumption 
that electrons are controlled perfectly.
Thus, in the current realistic situation, 
the controllability of countable electrons by electrodes 
would be the first issue to be addressed in the experiments.

\section{Conclusion}\label{sec:conclusion}
We have reviewed prospects for the QAM 
based on semiconductor nano-structures from the viewpoint of the integration of qubits.
In order to increase the coherence of qubits, the separation of qubits 
from their environment is important.
On the other hand, separation of qubits from their environment leads to 
weak control of the qubits. 
Thus, there is a trade-off in the relationship between qubit system with 
their environments, and here we focus on the charge qubit from 
the viewpoint of their natural mutual capacitive couplings.
In addition, it is considered that an integration of qubits is more difficult than 
integration of conventional CMOS transistors,
in particular when the qubit structure is quite different from 
conventional commercial semiconductor devices.
Thus, the fastest way, which also means the most economical way, is to 
use the current fabrication technologies to build a qubit system.
Accordingly, we have proposed a charge-qubit system 
using NAND flash memory. 
We also proposed similar qubit system based on a CQD system.
Because fabrication lines in the factory already exist, 
we hope business judgment will proceed experiments for testing our proposals.

\section*{Acknowledgments}
We thank T. Hiraoka, T. Hioki, T. Marukame, H. Goto, M. Hayashi, K. Kuboki and T. Otsuka for discussions.


\end{document}